\documentclass[12pt]{iopart}

\usepackage{graphics}
\begin{document}

\title{On the black hole limit of electrically counterpoised dust
configurations}

\author{Reinhard Meinel and Moritz H\"utten}

\address{Theoretisch-Physikalisches Institut,
University of Jena,\\
 Max-Wien-Platz 1, 07743 Jena, Germany}

\ead{meinel@tpi.uni-jena.de}
\begin{abstract}
By means of a simple scaling transformation any asymptotically flat
Papapetrou-Majumdar solution of the Einstein-Maxwell equations corresponding to
a localized regular distribution of electrically counterpoised dust can
be reformulated as a one-parameter family of solutions admitting a black hole
limit. In the limit, a characteristic separation of spacetimes occurs: From the
exterior point of view, the extreme Reissner-Nordstr\"om metric outside the
event horizon is formed. From the interior point of view, a regular,
non-asymptotically flat (and in general non-spherically symmetric) spacetime 
with the extreme Reissner-Nordstr\"om
near-horizon geometry at spatial infinity results.  
\end{abstract}


\vspace{2pc}

\section{Introduction}
It is well-known that a parametric sequence of spherically symmetric perfect
fluid bodies in equilibrium cannot come arbitrarily close to a black hole
state: The radius, in Schwarzschild coordinates, has to be greater than $9/8$
times the corresponding Schwarzschild radius $2M$ \cite{buchdahl}. In contrast,
some {\it rotating} fluid bodies in equilibrium admit a continuous black hole
limit. This was first demonstrated by Bardeen and Wagoner \cite{bw71} in the
limiting case of an infinitesimally thin disc of dust. A rigorous proof was
provided by Neugebauer and Meinel \cite{nm95} with the exact solution to the
disc problem, see also \cite{m02,rfe,klm}. 
In the limit, a ``separation of spacetimes''
occurs, leading to a non-asymptotically flat ``inner world'' and the extreme
Kerr spacetime outside the horizon as the ``outer world''. Numerically, such a
black hole limit was found for fluid rings as well \cite{akm} and it was proved
that it always leads to the {\it extremal} Kerr black hole \cite{m06}. Similar
phenomena were observed for limiting solutions to the static
Einstein-Yang-Mills-Higgs equations \cite{bfm,lw00}. However, the simplest
examples can be constructed by means of solutions to the static Einstein-Maxwell
equations belonging to the Papapetrou-Majumdar class \cite{p47,m47}, sometimes
called ``Bonnor stars''. This was demonstrated in
\cite{bw75,bonnor98,bonnor99,bonnor10,lw04,lz07}. The aim of the present paper
is to treat the general black hole limit of such ``electrically counterpoised
dust'' (ECD) configurations in much the same way as it was discussed for the
rotating
disc case in \cite{bw71}. It is shown that a characteristic separation of
spacetimes always occurs: From the exterior point of view, the extreme
Reissner-Nordstr\"om metric outside the event horizon is formed. From the
interior point of view, a regular, non-asymptotically flat spacetime with the
extreme Reissner-Nordstr\"om near-horizon geometry at spatial infinity results. 
Whereas the outer world is spherically symmetric, the inner world does not 
need to show any spatial symmetry in general. 
As a concrete example, we consider the black hole limit of an ECD
distribution of a triaxial ellipsoidal shape.

\section{Configurations of electrically counterpoised dust}
The Papapetrou-Majumdar class \cite{p47,m47} of static solutions to the 
Einstein-Maxwell equations can be described by the line element\footnote{We use
units 
in which the speed of light as well as Newton’s gravitational constant are equal
to 1 combined with the Gauss system for the electromagnetic quantities.} 
\begin{equation}
{\rm d}s^2=S^2({\rm d}x^2+{\rm d}y^2+{\rm d}z^2)-S^{-2}{\rm d}t^2.
\end{equation}
The corresponding energy-momentum tensor is given by
\begin{equation}
T_{ik}=\rho u_i u_k + T^{(\rm em)}_{ik}, \quad u^i=\delta^i_4\, S, \quad \rho
\ge 0, 
\quad S>0
\end{equation}
with
\begin{equation}
T^{(\rm em)}_{ik}=\frac{1}{4\pi}(F_{ij}{F_k}^j - \frac{1}{4}F^{mn}F_{mn}g_{ik}),
\quad F_{ik}=A_{k,i}-A_{i,k}
\end{equation}
and
\begin{equation}
A_i=-\delta^4_i\, \phi, \quad \phi=-\epsilon (S^{-1}-1), \quad \epsilon=\pm 1.
\end{equation}
The Einstein-Maxwell equations
\begin{equation}
R_{ik}-\frac{1}{2}R\,g_{ik}=8\pi T_{ik}, {\quad F^{ik}}_{;k}=4\pi J^i
\end{equation}
for the ECD case with
\begin{equation}
J^i=\sigma u^i,
\quad \sigma=\epsilon \rho
\end{equation}
($\sigma$ being the charge density) are equivalent to
\begin{equation}\label{eqS}
\Delta S\equiv \frac{\partial^2 S}{\partial x^2}+\frac{\partial^2 S}
{\partial y^2}+\frac{\partial^2 S}{\partial z^2}=-4\pi S^3\rho.
\end{equation}
Expressing $S$ and $\rho$ by means of two new functions $V$ and $\mu$ according
to 
\begin{equation}
S=1-V, \quad \rho=\frac{\mu}{S^3}\, ,
\end{equation}
\eref{eqS} can be rewritten as a Poisson equation: 
\begin{equation}
\Delta V=4\pi\mu.
\end{equation}
Asymptotically flat solutions, with a localized ECD distribution, can therefore 
be represented as a Poisson integral:
\begin{equation}\label{poisson}
V=-\int\frac{\mu({\bf r'})\,{\rm d}^3{\bf r'}}{|{\bf r}-{\bf r'}|}, \quad 
{\bf r}=(x,y,z).
\end{equation}
The asymptotic behaviour is given by
\begin{equation}
r \equiv |{\bf r}| \to\infty: \quad V\to -\frac{M}{r}, \quad g_{44}=-S^{-2} \to
-\left(1-\frac{2M}{r}\right) 
\end{equation}
with
\begin{equation}\label{eqM}
M=\int\mu({\bf r})\,{\rm d}^3{\bf r}
\end{equation}
being the gravitational mass, which is equal to the absolute value of the total charge 
$Q=\int\sigma S^3\,{\rm d}^3{\bf r}$.\footnote{Note that the gravitational binding energy
compensates the electromagnetic field energy for ECD configurations.}

\section{The black hole limit}
Any given solution belonging to an ECD distribution of finite extent 
can be characterized by a function $f({\bf r})$ vanishing outside some 
coordinate sphere of radius $r=R$ according to
\begin{equation}
\mu({\bf r})=f({\bf r})
\end{equation}
with
\begin{equation}\label{finite}
f({\bf r}) \equiv 0 \quad \mbox{for} \quad r>R.
\end{equation}
Alternatively, one can consider ECD distributions of infinite extent with a
function 
$f({\bf r})$ decaying sufficiently rapidly as $r\to\infty$.

Now, let us consider a corresponding one-parameter family of solutions 
char\-acterized by
\begin{equation}\label{family}
\mu({\bf r})=\alpha^3 f(\alpha {\bf r}), \quad \alpha > 0.
\end{equation}
For a function $f$ satisfying (\ref{finite}) this means
\begin{equation}\label{finite_a}
\mu({\bf r}) \equiv 0 \quad \mbox{for} \quad r>\frac{R}{\alpha}.
\end{equation}
Note that $M$ does not depend on the parameter $\alpha$.

For sufficiently small values of $\alpha$ the Newtonian limit ($|V|\ll 1$)
results. The black hole limit is obtained as
\begin{equation}
\alpha\to\infty.
\end{equation}
In the coordinates used so far the ECD distribution shrinks to the 
point $r=0$ in the black hole limit. The metric for $r>0$ describes 
the ``outer world'', while finite values of 
$\tilde r\equiv \alpha\, r$ reveal the ``inner world'' ($r=0$) 
in the limit $\alpha\to\infty$.

\subsection{The exterior point of view}
Formally, the limit $\alpha\to\infty$ leads to
\begin{equation}
\mu({\bf r})=M\delta({\bf r}).
\end{equation}
Accordingly, we get
\begin{equation}
r > 0: \quad V=-\frac{M}{r}, \quad S=1+\frac{M}{r}
\end{equation}
leading to
\begin{equation}\label{RN}
{\rm d}s^2=\left(1+\frac{M}{r}\right)^2({\rm d}x^2+{\rm d}y^2+
{\rm d}z^2)-\left(1+\frac{M}{r}\right)^{-2}{\rm d}t^2,
\end{equation}
i.e.\ the metric of an extremal Reissner-Nordstr\"om black hole outside the
event horizon, which is situated at $r=0$ in the isotropic coordinates used
here.\footnote{Note that the radial Schwarzschild coordinate $r_{\rm S}$ is given
by $r_{\rm S}=r+M$.}

\subsection{The interior point of view}
A completely different limit of the spacetime is obtained if the limit
$\alpha\to\infty$ is performed {\it after} the coordinate transformation
\begin{equation}\label{trafo}
\tilde x = \alpha x,\, \tilde y = \alpha y,\, \tilde z = \alpha z,\, 
\tilde t = \alpha^{-1} t
\end{equation}
leading to the line element
\begin{equation}\label{inner}
{\rm d}s^2={\tilde S}^2({\rm d}{\tilde x}^2+{\rm d}{\tilde y}^2+
{\rm d}{\tilde z}^2)-{\tilde S}^{-2}{\rm d}{\tilde t}^2
\end{equation}
with $\tilde S=S/\alpha$. In the limit $\alpha\to\infty$ this means
\begin{equation}
\tilde S = \left.\alpha^{-1} S\right|_{\alpha \to \infty} = 
\left.\alpha^{-1}(1-V)\right|_{\alpha \to \infty} =
\left.-\alpha^{-1}V\right|_{\alpha \to \infty}.
\end{equation}
Together with (\ref{poisson}), (\ref{family}) and (\ref{trafo}) we get
\begin{equation}\label{Stilde}
\tilde S = 
\int\frac{f(\tilde{\bf r}')\,{\rm d}^3\tilde{\bf r}'}
{|\tilde{\bf r}-\tilde{\bf r}'|}, 
\quad \tilde{\bf r}=(\tilde x,\tilde y,\tilde z).
\end{equation}
Note that all finite values of $\tilde r\equiv |\tilde{\bf r}|$ correspond to
$r=0$ in the limit! The ``inner world'' described by (\ref{inner}),
(\ref{Stilde}) is not asymptotically flat: Because of
\begin{equation}\label{asympt}
\tilde r\to\infty: \quad \tilde S\to\frac{M}{\tilde r}
\end{equation} 
the line element, for large $\tilde r$, approaches asymptotically
\begin{equation}\label{nhg}
{\rm d}s^2=\frac{M^2}{\tilde r^2}({\rm d}{\tilde x}^2+{\rm d}{\tilde y}^2+
{\rm d}{\tilde z}^2)-\frac{\tilde r^2}{M^2}{\rm d}{\tilde t}^2,
\end{equation}
which is the extreme Reissner-Nordstr\"om ``near-horizon geometry'' [resulting 
from (\ref{RN}) for small $r$], also known
as the Bertotti-Robinson metric or $AdS_2\times S^2$ spacetime. In fact,
this solution to the Einstein-Maxwell equations already appeared in a classic
paper by Levi-Civita that has recently been republished, see \cite{mm,lc}.
Note that in the special case of {\it finite} and {\it spherically symmetric}
ECD distributions the near-horizon geometry \eref{nhg} holds exactly for all
$\tilde r > R$.

The distribution of the mass-density $\rho$ as a function of $\tilde {\bf
r}$ remains regular in the limit
$\alpha\to\infty$ since 
\begin{equation}
\rho=\frac{\mu}{S^3}=\frac{f(\tilde {\bf r})}{\tilde S^3}
\end{equation}
for all $\alpha$.
Remember that $f(\tilde {\bf r})$ vanishes for
$\tilde r > R$ or decays sufficiently rapidly as $\tilde r\to\infty$.

\subsection{An example}
As a simple but nontrivial example we choose the function
\begin{equation}
f({\bf r})=\left\{
\begin{array}{l} \mu_0 \quad \mbox{for} \quad 
\frac{x^2}{a^2}+\frac{y^2}{b^2}+\frac{z^2}{c^2}\le 1\\
0 \quad \mbox{elsewhere}
\end{array} 
\right.
\end{equation}
\begin{equation}
\mbox{with} \quad \mu_0={\rm constant}.
\end{equation}
This represents a special ECD configuration with a triaxial ellipsoidal shape
($a$, $b$ and $c$ are 
the coordinate semiaxes).\footnote{Assuming $a\ge b\ge c$, $f({\bf r})$
satisfies \eref{finite} with $R=a$.} 
This example already shows the interesting effect of a non-spherically 
symmetric ``inner world'' mentioned in the introduction. For $a=b=c$ it 
reduces to the spherically symmetric model considered by Bonnor and 
Wickramasuriya \cite{bw75}.
The gravitational mass calculated according to \eref{eqM} is 
\begin{equation}
M=\frac{4\pi}{3}abc\, \mu_0. 
\end{equation}
Using the well-known formulae for the classical potential of a homogeneous
ellipsoid, see, e.g., \cite{h35}, the function $\tilde S$ given by \eref{Stilde}
can be expressed as
\begin{equation*}
\tilde S=\frac{3M}{4}\int\limits_{\lambda_0}^\infty 
\left[1 - \frac{\tilde x^2}{a^2+\lambda} - \frac{\tilde y^2}{b^2+\lambda} 
- \frac{\tilde z^2}{c^2+\lambda} \right]\frac{{\rm d}\lambda}
{\sqrt{(a^2+\lambda)(b^2+\lambda)(c^2+\lambda)}}\, ,
\end{equation*}
where the lower integration limit $\lambda_0$ is defined according to
\begin{equation}
\lambda_0=0 \quad \mbox{for} \quad 
\frac{\tilde x^2}{a^2}+\frac{\tilde y^2}{b^2}+\frac{\tilde z^2}{c^2}\le 1 
\quad \mbox{and}
\end{equation}
\begin{equation*}
\frac{\tilde x^2}{a^2+\lambda_0} + \frac{\tilde y^2}{b^2+\lambda_0} 
+ \frac{\tilde z^2}{c^2+\lambda_0} = 1 \quad \mbox{elsewhere}. 
\end{equation*}
%%%%%%%%%%%%%%%%%%%%%%%%%%%%%%%%%%%%%%%%%%%%%%%%%%%%%%%%%%%%%%%%%
\begin{figure}
\vspace{0.5cm}
\begin{center}
\scalebox{0.65}{\includegraphics{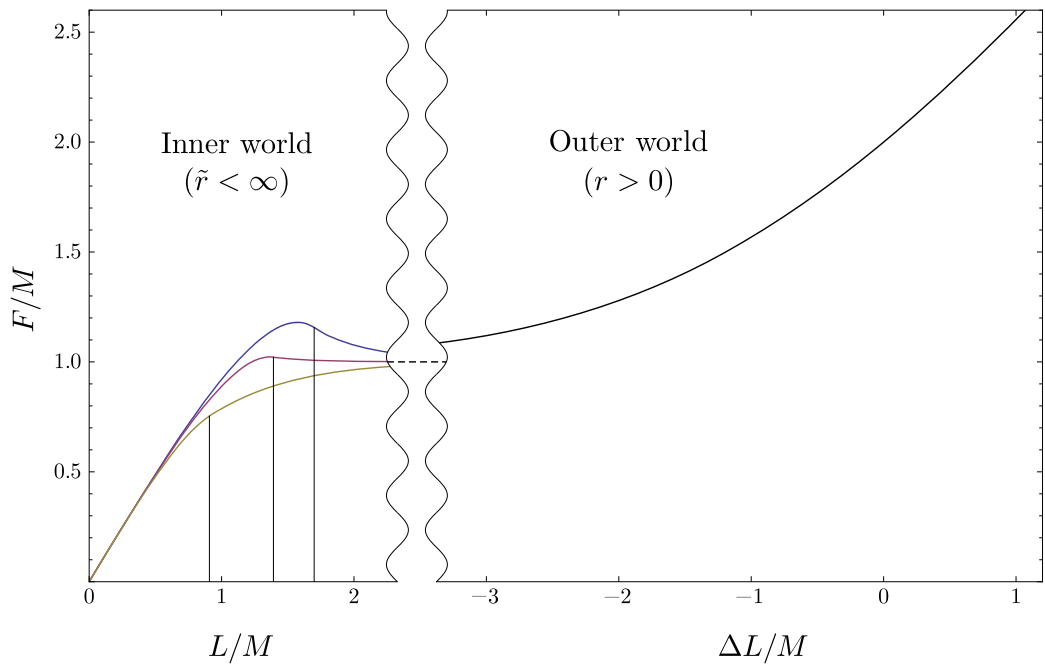}}
\end{center}
\caption{The metric function $F=rS=\tilde r\tilde S$ is plotted in the limit
$\alpha\to\infty$. In
the ``inner world'', the variation of $F$ in the direction of the three
principal axes of the ellipsoid is shown. $L$ is the proper distance from the
center. The vertical lines indicate the intersection points of the axes with the
surface of the ellipsoid. In the ``outer world'', $\Delta L$ denotes the proper
radial distance from the sphere $r=M$. The dashed line in the middle indicates
the asymptotic value $F=M$ which holds for $\tilde r\to \infty$ ($L\to\infty$)
as well as for $r\to 0$ ($\Delta L\to -\infty$), and corresponds to the
``near-horizon geometry''.}
\label{fig}
\end{figure}
%%%%%%%%%%%%%%%%%%%%%%%%%%%%%%%%%%%%%%%%%%%%%%%%%%%%%%%%%%%%%%%%%
Figure \ref{fig} combines the interior and the exterior point of view. 
We have chosen $b=0.8\,a$ and $c=0.5\,a$. 
The left part of the figure belonging to the ``inner world'' shows the 
variation of the function 
\begin{equation}\label{F}
F\equiv \tilde r\tilde S
\end{equation}
with the proper distance 
\begin{equation}
L=\int\limits_0^{\tilde r}\tilde S(\tilde r')\,{\rm d}\tilde r'
\end{equation}
from the centre of the ellipsoid\footnote{Strictly speaking, we mean a
coordinate ellipsoid.} along the $\tilde x$-, $\tilde y$- and $\tilde
z$-axis.
According to \eref{asympt}, the asymptotic value of $F$ is given by
\begin{equation}
\tilde r\to\infty: \quad F\to M
\end{equation}
in all directions. The right part of the figure belonging to the ``outer
world'' 
shows the variation of the function\footnote{The use of the same symbol for this
function as in \eref{F} is justified since $\tilde r\tilde S=rS$ for all finite
values of the parameter $\alpha$. In the strict limit $\alpha\to\infty$ the
inner and the outer world correspond to different limits of the spacetime, however.}
\begin{equation}
F\equiv rS
\end{equation}
in any radial direction (note that the outer world is spherically symmetric).
Here 
\begin{equation}
\Delta L=\int\limits_M^r S(r')\, {\rm d} r'
\end{equation}
denotes the proper radial distance from the arbitrarily chosen reference sphere
$r=M$ (or $r_{\rm S}=2M$ in Schwarzschild coordinates). The proper
radial distance of any place in the outer world ($r>0$) from $r=0$ (i.e.\
$r_{\rm S}=M$) is infinite. This is a characteristic feature of the extreme
Reissner-Nordstr\"om metric \eref{RN}. The limiting value of $F$ as $r\to 0$ is
given by
\begin{equation}
r\to 0: \quad F\to M.
\end{equation}
Thus the middle part of the figure with $F\equiv M$ (dashed line) belonging to
the near-horizon geometry
\eref{nhg} represents the ``connection'' between the inner
and the outer world. It should be stressed, however, that the inner world is a 
geodesically complete spacetime. 
These findings are in perfect analogy with the properties of the
extreme relativistic limit of the rotating disc as discussed in \cite{bw71}, see
figure 13 in that paper. 
 
\section{Discussion}
The one-parameter families of solutions considered in this paper
provide quasistatic routes from the Newtonian limit ($\alpha$ very small) up
to the black hole limit ($\alpha\to\infty$) where the characteristic
``separation of spacetimes'' occurs. Configurations close to the limit ($\alpha$
very large but not infinite) are still characterized by a comprehensive,
regular and asymptotically flat spacetime, which, for a distant exterior
observer, is almost indistinguishable from the spacetime of an extremal
Reissner-Nordstr\"om black hole outside the horizon. The term ``quasi-black
holes'' used by Lemos and others, see \cite{lz07}, is a nice denotation for
them. An interesting problem is the investigation of the stability of these
quasi-black holes. It is to be expected that small perturbations, e.g.\ the
infall of a small amount of (neutral) matter, will lead to a collapse and the
formation of a genuine, slightly subextremal black hole. The investigation of
the precise conditions for preventing the formation of naked singularities may
lead to new insights concerning ``cosmic censorship''.
In addition, we mention that the extremal Reissner-Nordstr\"om black holes
can be considered, in a sense, as classical models of point charges \cite{adm}. 

%\ack

\end{document}